\documentclass[10pt,final,notitlepage]{iopart}
\usepackage{graphicx}
\usepackage{epstopdf}
\usepackage{color}
\usepackage[usenames,dvipsnames]{xcolor}
\usepackage{dcolumn}% Align table columns on decimal point
\usepackage{bm}% bold math
\usepackage[colorlinks=true,linkcolor=Magenta,citecolor=Magenta]{hyperref}
\expandafter\let\csname equation*\endcsname\relax
\expandafter\let\csname endequation*\endcsname\relax
\usepackage{amsmath}
\usepackage{amssymb}
\usepackage{amsthm}

\def\rcurs{{\mbox{$\resizebox{.16in}{.08in}{\includegraphics{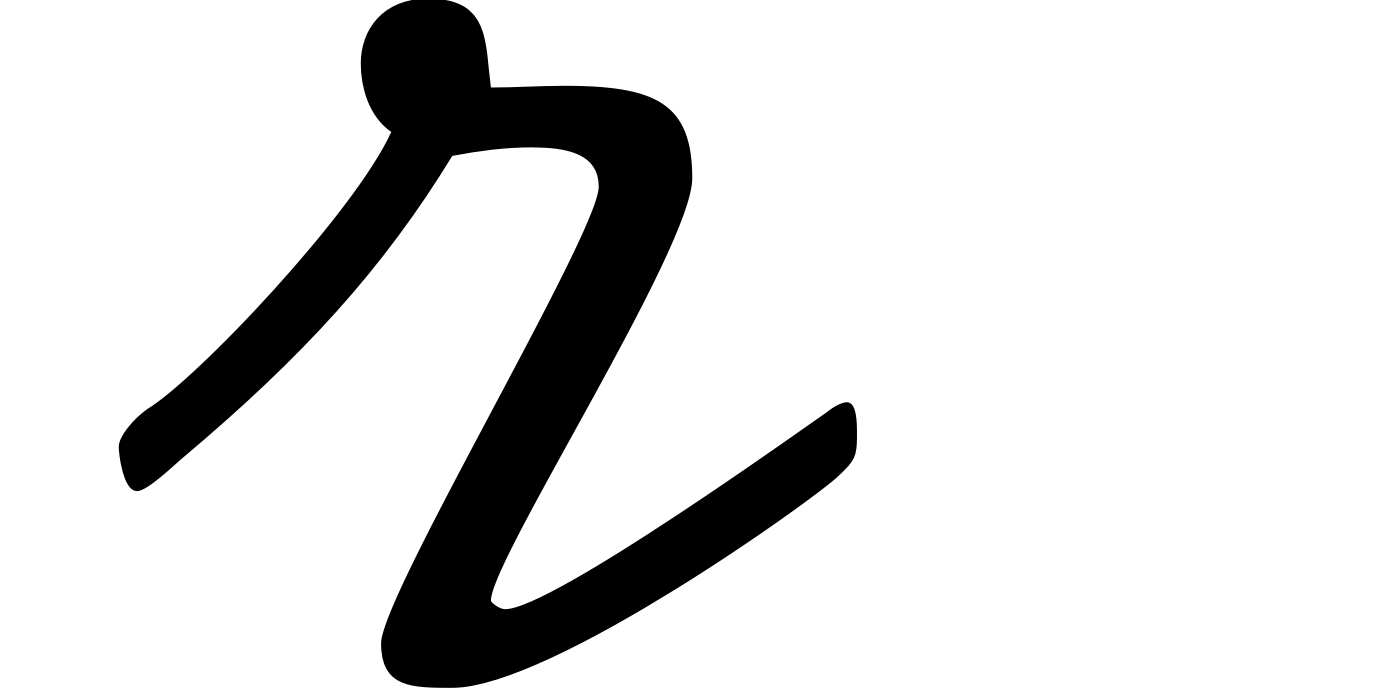}}$}}}

\begin{document}
\setlength{\abovedisplayskip}{3pt}
\setlength{\belowdisplayskip}{3pt}
\title[Extreme quantum nonequilibrium]{Extreme quantum nonequilibrium, nodes, vorticity, drift, and relaxation retarding states}
\author{Nicolas G Underwood$^{\hyperlink{first address}{1},\hyperlink{second address}{2}}$}
\address{$^{\hypertarget{first address}{1}}$Perimeter Institute for Theoretical Physics, 31 Caroline Street North, Waterloo, Ontario, Canada, N2L 2Y5}
\address{$^{\hypertarget{second address}{2}}$Kinard Laboratory, Clemson University, Clemson, South Carolina, USA, 29634}
\ead{nunderw@clemson.edu}
\begin{abstract}
Consideration is given to the behaviour of de Broglie trajectories that are separated from the bulk of the Born distribution with a view to describing the quantum relaxation properties of more `extreme' forms of quantum nonequilibrium. For the 2-dimensional isotropic harmonic oscillator, through the construction of what is termed the `drift field', a description is given of a general mechanism that causes the relaxation of `extreme' quantum nonequilibrium. Quantum states are found which do not feature this mechanism, so that relaxation may be severely delayed or possibly may not take place at all. A method by which these states may be identified, classified and calculated is given in terms of the properties of the nodes of the state. Properties of the nodes that enable this classification are described for the first time. 
\end{abstract}

\section{Introduction}
\renewcommand*{\thefootnote}{\fnsymbol{footnote}}
\renewcommand*{\thefootnote}{\arabic{footnote}}
\setcounter{footnote}{0}
De Broglie-Bohm theory \cite{deB28,B52a,B52b,Holl93} is the archetypal member of a class \cite{CS10,W07,DG97,Uthesis} of interpretations of quantum mechanics that feature a mechanism by which quantum probabilities arise dynamically \cite{CS10,AV91a,AV92,VW05,TRV12,ACV14,SC12} from standard ignorance-type probabilities. 
This mechanism, called `quantum relaxation', occurs spontaneously through a process that is directly analogous to the classical relaxation (from statistical nonequilibrium to statistical equilibrium) of a simple isolated mechanical system as described by the second thermodynamical law.
Demonstration of the validity of quantum relaxation \cite{VW05,EC06,CS10,SC12,TRV12,ACV14} has meant that the need to postulate agreement with quantum probabilities outright may be dispensed with. It has also prompted the consideration of `quantum nonequilibrium'--that is, violations of standard quantum probabilities, which in principle could be observed experimentally\footnote{This conclusion is not entirely without its critics. For an alternative viewpoint see references \cite{DGZ92,DT09}. For counterarguments to this viewpoint see references \cite{AV96,AV01}.}.
In other words, these theories allow for arbitrary nonequilibrium probabilities, only reproducing the predictions of standard quantum theory in their state of maximum entropy.
It has been conjectured that the universe could have begun in a state of quantum nonequilibrium \cite{AV91a,AV92,AV91b,AV96,AV01,UV15} which has subsequently mostly degraded, or that exotic gravitational effects may even generate nonequilibrium \cite{AV10,AV07,AV04b}.
If such nonequilibrium distributions were indeed proven to exist, this would not only demonstrate the need to re-assess the current quantum formalism, but could also generate new phenomena \cite{AV91b,AV08,AV02,PV06}, potentially opening up a large field of investigation.
In recent years some authors have focused their attention upon the prospect of measuring such violations of quantum theory. This work is becoming well developed in some areas, for instance with regard to measurable effects upon the cosmic microwave background \cite{AV10,CV13,CV15,CV16}. 
In other areas, for instance regarding nonequilibrium signatures in the spectra of relic particles \cite{UV15,UV16}, the literature is still in the early stages of development.

Other authors have instead focused upon the de Broglie trajectories and what this might tell us about the process of quantum relaxation.
In this regard, some statements may be made with certainty.
De Broglie trajectories tend to be chaotic in character \cite{F97,WP05,EC06,EKC07,TCE16,WS99}, and this is generally \cite{F97,WS99,WPB06}, but not always \cite{SC12}, attributed to the nodes of the wave function.
This chaos, in turn, is generally seen as the driving factor in quantum relaxation \cite{WPB06,TRV12,ECT17}.
Certainly, for relatively small systems with a small superposition and relatively minor deviations from equilibrium, relaxation takes place remarkably quickly \cite{VW05,TRV12,CS10}. The speed of the relaxation appears to scale with the complexity of the superposition \cite{TRV12}, whilst not occurring at all for non-degenerate energy states. 
Of course, the particles accessible to common experimentation have had a long and turbulent astrophysical past, and hence will have had ample time to relax. Consequently, it is not unreasonable to expect a high degree of conformance to standard quantum probabilities in everyday experiments.
This of course lends credence to the notion of quantum nonequilibrium despite the apparent lack of experimental evidence to date.
That said, there are still many open questions regarding the generality of quantum relaxation, and for sufficiently minimally interacting systems it may be possible that there are some windows of opportunity for quantum nonequilibrium to persist. The intention of this work is to address two such questions. These may be phrased as follows.\\
\\
\emph{If a system is very out-of-equilibrium, how does this affect the relaxation time?}\\
To date, every study of quantum relaxation has been concerned only with nonequilibrium distributions that begin within the bulk of the standard quantum probability distribution (variously referred to as $|\psi|^2$, the Born distribution or quantum equilibrium). There is however no \emph{a priori} reason why the initial conditions shouldn't specify a distribution that is far from equilibrium. (We refer to nonequilibrium distributions that are appreciably separated from the bulk of the Born distribution as `extreme' quantum nonequilibrium.) In the early development of the theory the main priority was to prove the validity of the process of relaxation, and so it was natural to choose initial distributions that had the possibility of relaxing quickly enough to be computationally tractable. When considering the possibility of quantum nonequilibrium surviving to this day however, this choice may appear as something of a selection bias. Certainly, the chaotic nature of the trajectories can make longer term simulations impractical, and so it is not entirely surprising that such situations have yet to make it into the literature. For the system discussed here (the 2 dimensional isotropic oscillator), state of the art calculations (for instance \cite{ACV14,KV16}) achieve evolution timescales of approximately $10^2$ to $10^4$ wave function periods and manage this only with considerable computational expense. A new numerical methodology described in section \ref{sec:drift} (construction of the `drift field') allows this computational bottleneck to be avoided, providing the means to describe the behaviour of systems which relax on timescales that may be longer by many orders of magnitude. \\
\\
\\
\emph{Is it possible that for some systems, relaxation does not take place at all?}\\
As has already been mentioned, relaxation does not occur for non-degenerate energy eigenstates. This is widely known and follows trivially as a result of the velocity field vanishing for such states. It is not yet clear, however, whether relaxation may be frozen or impeded more commonly and for more general states\footnote{Of course, since de Broglie-Bohm is time-reversible it is always possible to contrive initial conditions such that distributions appear to fall out of equilibrium. (A so-called `conspiracy' of initial conditions leading to `unlikely' entropy-decreasing behaviour \cite{AV91a}.) Indeed, since the cause of quantum relaxation is in essence the same as the cause of classical statistical relaxation \cite{Uthesis}, this is also true in classical mechanics. The possibility of such behaviour is excluded from the discussion for the usual reasons (see for instance reference \cite{D77}).}. Some authors have recently made some inroads regarding this question. One recent study \cite{KV16} found that for states that were perturbatively close to the ground state, trajectories may be confined to individual subregions of the state space, a seeming barrier to total relaxation. Another found that for some sufficiently simple systems there could remain a residual nonequilibrium that is unable to decay \cite{ACV14}. In this paper, a number of points are added to this discussion. Firstly, a general mechanism is identified that causes distributions that are separated from the bulk of the Born distribution to migrate into the bulk, a necessary precursor to relaxation for such systems. Also identified are states for which this mechanism is conspicuously absent. It is argued that for such states, the relaxation of systems separated from the bulk of $|\psi|^2$ will be at least severely retarded and possibly may not take place at all. It is shown that quantum states may be cleanly categorised according to the vorticity of their velocity field and that this vorticity may be calculated from the state parameters. With supporting numerical evidence, it is conjectured that states belonging to one of these categories always feature the relaxation mechanism, and also that for states belonging to another of the categories the mechanism is always absent. This provides a method by which states that feature either efficient or retarded (possibly absent) relaxation may be calculated. 

The article is structured as follows. Section \ref{sec:nodes} provides a description of the quantum system under investigation and provides an explanation of the role of nodes in quantum relaxation. A number of previously unknown properties of nodes are described which are valid under two general assumptions. These properties allow the categorisation of quantum states by their total vorticity. In section \ref{sec:drift} the `drift field' is introduced, its construction is outlined, and categories of drift field are described. An account is given of the mechanism by which `extreme' quantum nonequilibrium relaxes and it is argued that states that do not feature this mechanism will in the least exhibit severely retarded relaxation. A systematic study of 400 randomised quantum states is described which allows a link to be made between vorticity categories of section \ref{sec:nodes} and the categories of drift fields. These links are then stated as formal conjectures and supported with data from a further 6000 quantum states.
Finally, in section \ref{sec:conclusion} the main results of the investigation are summarised and implications for the prospect of quantum nonequilibrium surviving to this day are commented upon, with a view to future research directions.

\section{The system, its nodes and the vorticity theorem}\label{sec:nodes}
In studies of quantum relaxation, the two-dimensional isotropic harmonic oscillator is often chosen as the subject of investigation.
This is partly as it may be shown to be mathematically equivalent to a single uncoupled mode of a real scalar field \cite{AV07,AV08,ACV14,CV15,UV15,KV16}, granting it physical significance in studies concerning cosmological inflation scenarios or relaxation in high energy phenomena (relevant to potential avenues for experimental discovery of quantum nonequilibrium).
From a more practical perspective, two-dimensional systems also lend themselves well to plotting, allowing a more illustrative description to be made.
There is yet to be a systematic study of relaxation in many dimensions, and such a study would certainly add to our current understanding of quantum relaxation. That said, one primary intention of this work is to introduce a method which is tailored to provide an effective description of more `extreme' forms of nonequilibrium, and for this purpose it is useful to have other works with which to draw comparison. 
For these reasons, the two-dimensional isotropic harmonic oscillator is the subject of this investigation.
It has been useful to depart from convention, however, by working primarily with an `angular' basis of states that is built, not from the energy states of two one-dimensional oscillators (what will be referred to as a Cartesian basis), but from states that are simultaneous eigenstates of total energy and angular momentum. This is for a number of reasons. Principally, many of the results described in this section have been found as a result of using this basis. Also, as described in section \ref{sec:drift}, the long-term drift of trajectories decomposes well into radial and angular components, the size of which may differ by orders of magnitude. The angular basis is naturally described in terms of polar coordinates, and so its use helps to reduce numerical errors. Finally, the angular basis is arguably more natural as it takes advantage of the symmetry of the system. There are many sets of Cartesian bases but only one angular basis. Adoption of a Cartesian basis has lead to the study of states that could be considered `unusual' or `finely-tuned'. For instance, in \cite{ACV14} the authors used the notation $M=4$ and $M=25$ to denote the `first 4' and the `first 25' energy states. By $M=4$ it was meant that the state had $\psi_{00}$, $\psi_{10}$, $\psi_{01}$ and $\psi_{11}$ components (using the notation below). It is the case, however, that $\psi_{11}$ has the same energy as $\psi_{20}$ and $\psi_{02}$. Furthermore, it may easily be shown that under rotations energy states mix amongst other states of equal energy. A rotation of such a state would therefore supplement the superposition with $\psi_{20}$ and $\psi_{02}$ parts, turning what was referred to as $M=4$ into a superposition of 6 states. Similarly, rotated axes transform what was referred to as $M=25$ into the lowest 45 energy states. It is debatable whether states that have been selected in this manner should or should not be considered finely-tuned. In the course of the present study, however, it was found that such states exhibit quite unusual, indeed pathological behaviour. It is the intention here to keep the discussion general and hence to avoid discussion of such states, which will be the subject of another paper. In order to exclude such `fine-tuning', the following criterion is held throughout this work. If any state may be considered to have been selected using criteria that restricts the Hilbert space in such a manner that the resulting subspace is measure-zero, then this shall be considered fine-tuning and worthy to be disregarded from discussion. For example, superpositions with components that are exactly the same magnitude or that differ in phase by exactly $\pi/2$ are considered to be finely-tuned. Of course, a state in a Hilbert space that is completely unrestricted in this manner would require an infinite number of parameters to specify and make the problem intractable. Hence, it is necessary to make one exception to the rule. This exception is chosen to be an upper limit to the allowable energy in a superposition that, following the notation of \cite{ACV14} will be specified by the symbol $M$. With the stated considerations, however, $M$ should be understood to take only the values 1, 3, 6, 10, 15 etc. Note that $M$ may be unambiguously referred to in this manner as for instance, accounting for degeneracy, a combination of the first 15 energy states remains a combination of the first 15 energy states regardless of the basis used (Cartesian, rotated Cartesian, or angular). 

It is useful to develop the Cartesian description in parallel with the angular description.
The subject of investigation is an isotropic two-dimensional harmonic oscillator of mass $m$ and frequency $\omega$.
The standard Cartesian coordinates $(q_x,q_y)$, radial coordinate $r$, and time $t$, are replaced with dimensionless counterparts $(Q_x,Q_y)$, $\eta$, and $T$ respectively. These are related by
\begin{align}
Q_x=\sqrt{\frac{m\omega}{\hbar}}q_x,\quad
Q_y=\sqrt{\frac{m\omega}{\hbar}}q_y,\quad
\eta=\sqrt{\frac{m\omega}{\hbar}}r,\quad
T=\omega t.
\end{align}
The symbol $\varphi$ is used to denote the anticlockwise angle from the $Q_x$ axis. The partial derivative with respect to dimensionless time $T$ is denoted by the placement of hollow dot $\overset{\circ}{}$ above the subject of the derivative. 
In these coordinates the Schr\"{o}dinger equation for the oscillator is
\begin{align}
\frac{1}{2}\left(\partial_{Q_x}^2+\partial_{Q_y}^2+Q_x^2+Q_y^2\right)\psi&=i\overset{\circ}{\psi},\\
\frac{1}{2}\left(\partial_\eta^2+\eta^{-1}\partial_\eta+\eta^{-2}\partial_\varphi^2+\eta^2\right)\psi&=i\overset{\circ}{\psi}.
\end{align}
From these equations, continuity equations
\begin{align}
0&=\overset{\circ}{|\psi|^2}+\partial_{Q_x}\left[|\psi|^2\text{Im}(\partial_{Q_x}\psi/\psi)\right]+\partial_{Q_y}\left[|\psi|^2\text{Im}(\partial_{Q_y}\psi/\psi)\right],\\
0&=\overset{\circ}{|\psi|^2}+\eta^{-1}\left[\partial_\eta\left(\eta|\psi|^2\text{Im}(\partial_\eta\psi/\psi)\right)+\partial_\varphi\left(|\psi|^2\eta^{-1}\text{Im}(\partial_\varphi\psi/\psi)\right)\right],
\end{align}
may be found in the usual manner, from which follow the de Broglie guidance equations
\begin{align}\label{xy_guidance}
\overset{\circ}{Q_x}=\text{Im}(\partial_{Q_x}\psi/\psi),&\quad\overset{\circ}{Q_y}=\text{Im}(\partial_{Q_y}\psi/\psi),\\
\overset{\circ}{\eta}=\text{Im}(\partial_\eta\psi/\psi),&\quad\overset{\circ}{\varphi}=\eta^{-2}\text{Im}(\partial_\varphi\psi/\psi).\label{angular_guidance}
\end{align}
The expansion in terms of Cartesian basis states $\psi_{n_x n_y}$ is expressed
\begin{align}\label{xy_state}
\psi=\sum_{n_n n_y}D_{n_x n_y}e^{-i(n_x+n_y)T}\psi_{n_x n_y}(Q_x,Q_y),
\end{align}
where the basis states are
\begin{align}
\psi_{n_x n_y}=\frac{1}{\sqrt{n_x!n_y!}}\left(a_x^\dagger\right)^{n_x}\left(a_y^\dagger\right)^{n_y}\psi_{00}=\frac{H_{n_x}(Q_x)H_{n_y}(Q_y)}{\sqrt{2^{n_x}2^{n_y}n_x!n_y!}}\psi_{00},\label{xy_basis}
\end{align}
$\psi_{00}$ is the ground state, and $a_x^\dagger$ and $a_y^\dagger$ denote the raising operators in their respective $Q_x$ and $Q_y$ directions. It will sometimes also be useful to express the complex coefficients $D_{n_x n_y}$ in terms of the real $d_{n_x n_y}$ and $\theta_{n_x n_y}$ which are related as $D_{n_x n_y}=d_{n_x n_y}\exp(i\theta_{n_x n_y})$. 

The angular basis is composed of simultaneous eigenstates of the total energy and angular momentum\footnote{Much of the notation adopted follows the conventions of \cite{CT}. In particular, it may be useful to note that the subscripts of symbols $n_d$ and $n_g$ derive from the French for right and left.}. The basis states $\chi_{n_d n_g}$ may be constructed in an analogous manner to equation \eqref{xy_basis}, as
\begin{align}\label{angular_basis}
\chi_{n_d n_g}=\frac{1}{\sqrt{n_d!n_g!}}(a_d^\dagger)^{n_d}(a_g^\dagger)^{n_g}\chi_{00}=e^{i(n_d-n_g)\varphi}f_{n_d n_g}(\eta)\chi_{00},
\end{align}
where $a_d^\dagger$ and $a_g^\dagger$ are right and left raising operators related to the usual Cartesian raising operators by
\begin{align}
a_d^\dagger=\frac{1}{\sqrt{2}}\left(a_x^\dagger + ia_y^\dagger\right),\quad a_g^\dagger=\frac{1}{\sqrt{2}}\left(a_x^\dagger-ia_y^\dagger\right).
\end{align}
The $f_{n_d n_g}(\eta)$ are polynomials of order $n_g+n_d$ in radial coordinate $\eta$, the first 15 of which are
\begin{align}\label{eq:f}
f_{00}=1,\quad f_{10}=f_{01}=\eta,\quad &f_{20}=f_{02}=\frac{1}{\sqrt{2}}\eta^2, \quad f_{11}=\eta^2-1,\nonumber\\
f_{30}=f_{03}=\frac{1}{\sqrt{6}}\eta^3,\quad &f_{21}=f_{12}=\frac{1}{\sqrt{2}}\left(\eta^3-2\eta\right),\\
f_{40}=f_{04}=\frac{1}{\sqrt{24}}\eta^4,\quad f_{31}=f_{13}=&\frac{1}{\sqrt{6}}\left(\eta^4-3\eta^2\right),\quad f_{22}\frac{1}{\sqrt{4}}\left(\eta^4-4\eta^2+2\right).\nonumber
\end{align}
The energy eigenvalue of state $\chi_{n_d n_g}$ is proportional to $n_d + n_g$ and its angular momentum eigenvalue is proportional to $n_d -n_g$. The state expansion in the angular basis states is denoted
\begin{align}\label{angular_expansion}
\psi=\sum_{n_d n_g}C_{n_d n_g}e^{-i(n_d+n_g)T}\chi_{n_d n_g},
\end{align}
and it is occasionally useful to express the complex coefficients in the polar form $C_{n_d n_g}=c_{n_d n_g}\exp(i\phi_{n_d n_g})$.
Components for up to $M=15$ states may be translated between bases using
\begin{align}\label{eq:coef_transforms}
&D_{00}=C_{00},\quad
D_{10}=\sqrt{1/2}C_{10}+\sqrt{1/2}C_{01},\quad
D_{01}=i\sqrt{1/2}C_{10}-i\sqrt{1/2}C_{01},\nonumber\\
&D_{20}=\sqrt{1/4}C_{20}+\sqrt{2/4}C_{11}+\sqrt{1/4}C_{02},\quad
D_{11}=i\sqrt{1/2}C_{20}-i\sqrt{1/2}C_{02},\nonumber\\
&D_{02}=-\sqrt{1/4}C_{20}+\sqrt{2/4}C_{11}-\sqrt{1/4}C_{02},\nonumber\\
&D_{30}=\sqrt{2/16}C_{30}+\sqrt{6/16}C_{21}+\sqrt{6/16}C_{12}+\sqrt{2/16}C_{03},\nonumber\\
&D_{21}=i\sqrt{6/16}C_{30}+i\sqrt{2/16}C_{21}-i\sqrt{2/16}C_{12}-i\sqrt{6/16}C_{03},\nonumber\\
&D_{12}=-\sqrt{6/16}C_{30}+\sqrt{2/16}C_{21}+\sqrt{2/16}C_{12}-\sqrt{6/16}C_{03},\\
&D_{30}=-i\sqrt{2/16}C_{30}+i\sqrt{6/16}C_{21}-i\sqrt{6/16}C_{12}+i\sqrt{2/16}C_{03},\nonumber\\
&D_{40}=\sqrt{1/16}C_{40}+\sqrt{4/16}C_{31}+\sqrt{6/16}C_{22}+\sqrt{4/16}C_{40}+\sqrt{1/16}C_{04},\nonumber\\
&D_{31}=i\sqrt{4/16}C_{40}+i\sqrt{4/16}C_{31}-i\sqrt{4/16}C_{13}-i\sqrt{4/16}C_{04},\nonumber\\
&D_{22}=-\sqrt{6/16}C_{40}+\sqrt{4/16}C_{22}-\sqrt{6/16}C_{04},\nonumber\\
&D_{13}=-i\sqrt{4/16}C_{40}+i\sqrt{4/16}C_{31}-i\sqrt{4/16}C_{13}+i\sqrt{4/16}C_{04},\nonumber\\
&D_{40}=\sqrt{1/16}C_{40}-\sqrt{4/16}C_{31}+\sqrt{6/16}C_{22}-\sqrt{4/16}C_{40}+\sqrt{1/16}C_{04}.\nonumber
\end{align}

The nodes of the wave function are often cited as the primary source of chaos in de Broglie trajectories \cite{F97,WS99,EKC07,ECT17}.
This chaos is in turn thought to be one the primary driving factors in quantum relaxation \cite{WPB06,TRV12,ECT17}.
As is described in the next section, the nodes (which are mostly to be found amongst the bulk of the Born distribution) also play an important role in the long-term relaxation of systems that may be far away and exhibit very regular behaviour. Indeed, global properties of the `drift field' described in section \ref{sec:drift} may be inferred from the properties of the nodes. It is therefore useful to know some of these properties. 
For the sake of clarity, these properties are listed, giving a short justification of each. For the sake of brevity and simplicity the system is taken to be the two-dimensional isotropic oscillator with the dual assumptions of no fine-tuning and limited energy expressed earlier, although most if not all of these properties may be easily generalised. (Properties (i) through (vi) are widely known or have been previously noted and are included to aid the reader. To our knowledge, properties (vii) through (xii) are new.) \\
\\
\textbf{(i) Nodes are points.} This follows trivially as $\psi=0$ places two real conditions on the two-dimensional space.\\
\textbf{(ii) The velocity field has zero vorticity away from nodes.} By writing the wave function in the complex exponential form $\psi=|\psi|\exp(iS)$, guidance equations \eqref{xy_guidance} may be seen to define a velocity field, $v=(\overset{\circ}{Q_x},\overset{\circ}{Q_y})=\nabla S$, that is the (two-dimensional) gradient of complex phase $S$. Consequently the vorticity (as the curl of the velocity $\nabla\times v$) must vanish everywhere except on nodes, where $S$ is ill-defined.\\
\textbf{(iii) The vorticity of nodes is `quantised'.} This observation was originally made by Dirac \cite{Dirac31} and is well known to the field \cite{Holl93}. It is also well known in other fields like chemical physics \cite{HCP74,HGB74,W06}. By Stokes' theorem, the line integral of the velocity field around a closed curve $\partial\Sigma$, that defines the boundary of some region $\Sigma$ which does not contain a node, must vanish; $\oint_{\partial\Sigma}v.\mathrm{d}l=\int_{\Sigma}\nabla\times v \mathrm{d}\Sigma=0$.
If on the other hand, the region $\Sigma$ is chosen such that it contains a node, this need not be the case.
The single-valuedness of $\psi$, however, assures that along any closed path, the phase $S$ can only change by some integer number of $2\pi$ from its initial value, i.e.
\begin{align}\label{eq:v_around_node}
\oint_{\partial \Sigma}v.\mathrm{d}l=\oint_{\partial \Sigma}\mathrm{d}S=2\pi n,
\end{align}
for integer $n$.
In this manner, it is said that the vorticity is `quantised'. Note that $\nabla\times v$ is actually ill-defined on the node and that the vorticity $\mathcal{V}$ of a node should instead be understood to refer to $\oint_{\partial \Sigma}v.\mathrm{d}l$ as evaluated around a closed path enclosing only the node in question.\\
\textbf{(iv) Nodes generate chaos.} It is possible to go into much detail on this point. For the intricacies regarding this process, articles \cite{EKC07,ECT17} and references therein are recommended. It is possible, however, to illustrate this point in the following simple way. 
 By taking the region $\Sigma$ in equation \eqref{eq:v_around_node} to be a ball $B_{\rcurs}$ of radius $\rcurs$, centred on some node, it may be concluded that the component of the velocity field $v$ around the node varies in proportion to $1/\rcurs$.
In other words, a trajectory that approaches close to a node will tend to circle the node with a speed that is strongly dependent upon how close the trajectory manages to get. Two trajectories that are initially near to each other and come close to a node may only differ by a small amount in their approach, but due to the $1/\rcurs$ dependence, this small difference may cause a significant difference in how the trajectories circle the node. The trajectories will likely be scattered in completely different directions.
This `butterfly effect' may lead to highly erratic trajectories and is commonly thought to be one of the primary causes of chaos in de Broglie trajectories.\\
\textbf{(v) Nodes pair create/annihilate.} This is explained for the three dimensional case in \cite{H77}. For our purposes, the condition of a node ($\psi=0$) may be regarded as two separate real conditions upon the 2-dimensional space. Each of these conditions defines a (plane algebraic) curve that evolves with time with nodes appearing where the curves intersect. Two curves that are not initially intersecting may begin to do so, creating a node. In an open system however, this must take place either at infinity (as for instance must be the case for two straight lines), or if the curves begin to intersect at a finite coordinate, they must necessarily intersect twice. This creates the appearance of the pair-creation of nodes. Similarly, two curves may cease to intersect, creating the appearance of annihilation. (Consider for instance an ellipse whose path crosses a static straight line. Upon initial contact, two nodes are created. These then annihilate when the ellipse completes its passage through the line.)\\
\textbf{(vi) Vorticity is locally conserved.} This was noted by \cite{WPB06} although to our knowledge there is yet to be an explanation of why this is the case. By local conservation it is meant that pair creation and annihilation of nodes may only take place between two nodes of opposite vorticity. Consider for instance the line integral $\oint_{\partial\Sigma}v.\mathrm{d}l=\oint_{\partial\Sigma}\mathrm{d}S$ around a region $\Sigma$ the instant before a pair of nodes is created. It is expected that $S$ is smooth in both space and time everywhere but on the node, and hence immediately after the pair creation there cannot be a jump to one of the quantised values of $2\pi n$ allowed by equation \eqref{eq:v_around_node}. Hence, it must be the case that the nodes created are equal and opposite in their vorticity. It follows that a local conservation law must hold, but as nodes may appear from or disappear to infinity in a finite time, this does not guarantee global conservation.\\
\textbf{(vii) Nodes have $\pm 2\pi$ vorticity.}
This stronger condition than property (iii) may be arrived at with the no fine-tuning assumption.
Suppose there exists a node in $\psi$ at coordinate $Q_{x_0},Q_{y_0}$. Since $\psi$ is analytic (in the real analysis sense), at any moment in time and in some small region around the node, it may be Taylor expanded
\begin{align}
\psi(Q_x,Q_y)&=a_x(Q_x-Q_{x_0})+a_y(Q_y-Q_{y_0})+\mathcal{O}\left[(Q_x-Q_{x_0})^2,(Q_y-Q_{y_0})^2\right]\nonumber\\
&=a_x\epsilon\cos\theta +a_y\epsilon\sin\theta+\mathcal{O}(\epsilon^2),
\end{align}
where $a_x$ and $a_y$ are complex constants that depend upon the state parameters, the coordinate of the node and the time. Vanishing $a_x$ or $a_y$ would either be instantaneous or require fine tuning of the state parameters. It is assumed therefore that $a_x$ and $a_y$ are non-zero. The $\epsilon$ and $\theta$ are polar coordinates centred upon the node. The vorticity of the node is calculated by integrating the change in phase around the edge of a small ball $B_{\epsilon}(Q_{x_0},Q_{y_0})$ centred on the node,
\begin{align}\label{eq:that_one}
\mathcal{V}=\oint_{\partial B_{\epsilon}(Q_{x_0},Q_{y_0})} \mathrm{d}S
=\oint_{\partial B_{\epsilon}(Q_{x_0},Q_{y_0})}\text{Im}\partial_\theta\log(a_x\epsilon\cos\theta+a_y\epsilon\sin\theta)\mathrm{d}\theta.
\end{align}
The radius of the ball $\epsilon$ is taken to be small enough firstly to ignore $\mathcal{O}(\epsilon^2)$ terms, but also to exclude other nodes from the interior of the ball. The factor $\epsilon$ then drops out of \eqref{eq:that_one} once the imaginary part is taken. By making the substitution $z=a_x\cos\theta+a_y\sin\theta$ the vorticity may be written
\begin{align}
\mathcal{V}=\text{Im}\oint_{\gamma(\theta)}\frac{\mathrm{d}z}{z}=\pm 2\pi,
\end{align}
where the final equality is found as the contour $\gamma(\theta)=a_x\cos\theta+a_y\sin\theta$ winds around $z=0$ once. The sign of the vorticity is determined by whether the path $\gamma$ is clockwise or anticlockwise, which may be found to be equal to the sign of $\sin(\text{Arg}(a_x)-\text{Arg}(a_y))$. Note that states which are finely-tuned in the manner described may not satisfy this property. For instance, the exact angular momentum state $\chi_{3\,0}$ has a pole of vorticity $6\pi$ at the origin.\\
\textbf{(viii) Vorticity is globally conserved.}
The total vorticity of a state may be written
\begin{align}\label{A1}
\mathcal{V}_\text{tot}=\lim_{\eta\rightarrow\infty}\oint_{\partial B_{\eta}(0)}\mathrm{d}S=\lim_{\eta\rightarrow\infty}\int_{0}^{2\pi}\frac{\partial S}{\partial \varphi}\,\mathrm{d}\varphi=\lim_{\eta\rightarrow\infty}\text{Im}\int_{0}^{2\pi}\frac{\partial_\varphi\psi}{\psi}\,\mathrm{d}\varphi.
\end{align}
By substituting \eqref{angular_basis} into \eqref{angular_expansion}, an arbitrary state may be written
\begin{align}\label{A2}
\psi&= \sum_{n_d n_g}C_{n_d n_g}\exp\left[-i(n_d+n_g)T+i(n_d-n_g)\varphi\right]f_{n_dn_g}(\eta)\chi_{00}(\eta),
\end{align}
and hence the integrand in \eqref{A1} becomes
\begin{align}\label{eq:dpsi}
\frac{\partial_\varphi\psi}{\psi}=\frac{\sum_{n_d n_g}i(n_d-n_g)C_{n_d n_g}\exp\left[-i(n_d+n_g)T+i(n_d-n_g)\varphi\right]f_{n_dn_g}(\eta)}{\sum_{n_d n_g}C_{n_d n_g}\exp\left[-i(n_d+n_g)T+i(n_d-n_g)\varphi\right]f_{n_dn_g}(\eta)}.
\end{align}
The polynomials $f_{n_d n_g}(\eta)$ are of order $n_d+n_g$ and so in the limit $\eta\rightarrow\infty$, only the highest order terms in the numerator and denominator expansions will contribute. Since these terms have identical time-phase factors $\exp[-i(n_d+n_g)T]$, these cancel leaving the integrand, and hence the total vorticity, time independent.\\
\textbf{(ix) States of total vorticity $\mathcal{V}_\text{tot}$ have a minimum of $\mathcal{V}_\text{tot}/2\pi$ nodes.} This follows from properties (vii) and (viii). Of course, pair creation allows there to be more than this number.\\
\textbf{(x) The `vorticity theorem' provides a simple method by which the total vorticity of a state may be calculated.}
The vorticity theorem may be stated as follows. \\
\emph{Let $\psi$ be a state of the 2-dimensional isotropic oscillator with an expansion \eqref{angular_expansion} that is bounded in energy by $n_d+n_g=m$. Let $f(z)$ be the complex Laurent polynomial,
\begin{align}\label{laurent_polynomial}
f(z)=\sum_{n_d=0}^{m}\frac{C_{n_d\, (m-n_d)}}{\sqrt{n_d!(m-n_d)!}}z^{2n_d-m},
\end{align}
formed from the coefficients corresponding to the states of highest energy in the expansion. If no zeros or poles of $f(z)$ lie on the complex unit circle $\partial B_1(0)$, then the total vorticity $\mathcal{V}_\text{tot}$ of the state $\psi$ is
\begin{align}\label{eq:tot_vort}
\mathcal{V}_\text{tot}&=2\pi \left[\#\text{ of zeros of }f(z)\text{ in }B_{1}(0)-\#\text{ of poles of }f(z)\text{ in }B_{1}(0)\right],
\end{align}
the difference between the number of zeros and poles of $f(z)$ in the unit disk $B_{1}(0)$ multiplied by $2\pi$, where it is understood that the counting should be according to multiplicity.}\\
This may be proven by inserting \eqref{eq:dpsi} into \eqref{A1}, assuming the energy upper bound $n_d+n_g=m$, and taking the limit $\lim_{\eta\rightarrow\infty}$. This results with the expression
\begin{align}\label{A4}
\mathcal{V}_\text{tot}&=\text{Im}\int_0^{2\pi}\frac{\sum_{n_d=0}^m i(2n_d-m)C_{n_d (m-n_d)}[n_d!(m-n_d)!]^{-\frac{1}{2}}\exp\left[i(2n_d-m)\varphi\right]}{\sum_{n_d=0}^mC_{n_d (m-n_d)}[n_d!(m-n_d)!]^{-\frac{1}{2}}\exp\left[i(2n_d-m)\varphi\right]}\mathrm{d}\varphi,
\end{align}
where the factors of $[n_d!(m-n_d)!]^{-\frac{1}{2}}$ come from the leading factors of $f_{n_d n_g}(\eta)$ (see equation \eqref{eq:f}). 
By writing $z=e^{i\varphi}$, this may be expressed as a complex contour integral around the unit circle $\partial B_{1}(0)$,
\begin{align}
\mathcal{V}_\text{tot}&=\oint_{\partial B_{1}(0)}\frac{\sum_{n_d=0}^m i(2n_d-m)C_{n_d (m-n_d)}[n_d!(m-n_d)!]^{-\frac{1}{2}}z^{2n_d-m}}{\sum_{n_d=0}^mC_{n_d (m-n_d)}[n_d!(m-n_d)!]^{-\frac{1}{2}}z^{2n_d-m}}(-iz^{-1})\mathrm{d}z\\
&=\text{Im}\oint_{\partial B_{1}(0)}\frac{f'(z)}{f(z)}\mathrm{d}z,
\end{align}
where $f(z)$ is defined as in \eqref{laurent_polynomial}. The Argument Principle of complex analysis (see for instance p.119 of \cite{Conway}) may then be used to arrive at the final result \eqref{eq:tot_vort}.\\
\\
\\
\textbf{(xi) A state that is bounded in energy by $n_d+n_g=m$ may only take the following $m+1$ possible total vorticities,}
\begin{align}\label{eq:poss_vort}
\mathcal{V}_\text{tot}=-2\pi m,\; -2\pi m+4\pi,\;\dots,\;2\pi m-4\pi,\;2\pi m.
\end{align}
This follows from taking a factor of $z^{-m}$ out of \eqref{laurent_polynomial}. The $z^{-m}$ contributes an $m$th order pole at the origin and the remaining polynomial has the property that if $z$ is a zero then so also is $-z$ prompting a rephrasing of the vorticity theorem as follows.\\
\emph{Let $\psi$ be a state of the 2-dimensional isotropic oscillator with an expansion \eqref{angular_expansion} that has an upper limit in energy, $n_d+n_g=m$. Let $g(z)$ be the complex polynomial,
\begin{align}\label{polynomial}
g(z)=\sum_{n_d=0}^{m}\frac{C_{n_d\, (m-n_d)}}{\sqrt{n_d!(m-n_d)!}}z^{n_d}.
\end{align}
formed from the coefficients corresponding to the states of highest energy in the expansion. If there are no zeros on the unit circle $\partial B_{1}(0)$, the total vorticity $\mathcal{V}_\text{tot}$ of the state $\psi$ is
\begin{align}
\mathcal{V}_\text{tot}&=2\pi \left[2\times\#\text{ of zeros of }g(z)\text{ in }B_{1}(0)-m\right],
\end{align}
where again it is understood that the counting should be according to multiplicity.}\\
An $m$th order complex polynomial has $m$ roots, each of which may or may not be inside the unit ball. Accounting for every possible case, it may be concluded that a state of maximum energy $m$ may have the $m+1$ possible total vorticities expressed in equation \eqref{eq:poss_vort}.\\
\textbf{(xii) A state that is bounded in energy by $n_d+n_g=m$ has a maximum of $m^2$ nodes.}
Since $\psi_{00}$ is everywhere non-zero, the condition of a node may be written
\begin{align}
\sum_{n_n n_y}D_{n_x n_y}e^{-i(n_x+n_y)T}\frac{H_{n_x}(Q_x)H_{n_y}(Q_y)}{\sqrt{2^{n_x}2^{n_y}n_x!n_y!}}=0,
\end{align}
the real and imaginary parts of which define plane algebraic curves of degree $m$. B\'{e}zout's theorem (originally to be found in Newton's Principia) states that the maximum number of times two such curves may intersect is given by the product of their degrees, in this case $m^2$.

\section{The drift field and its structure}\label{sec:drift}
\begin{figure}
\includegraphics{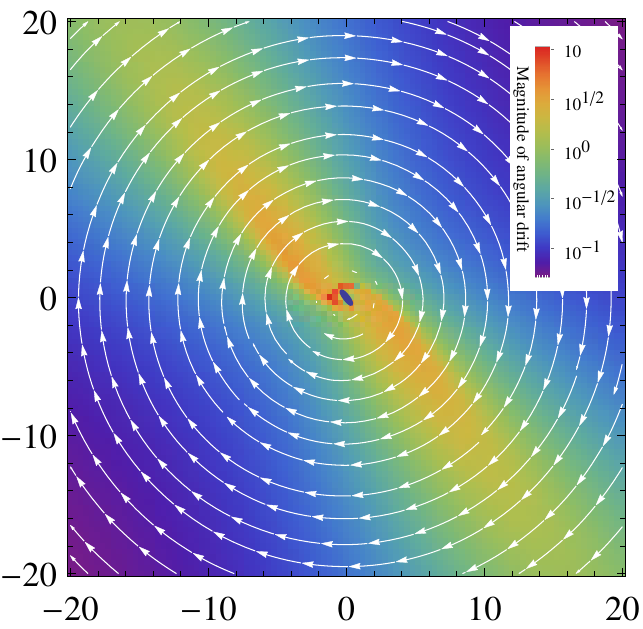}
\includegraphics{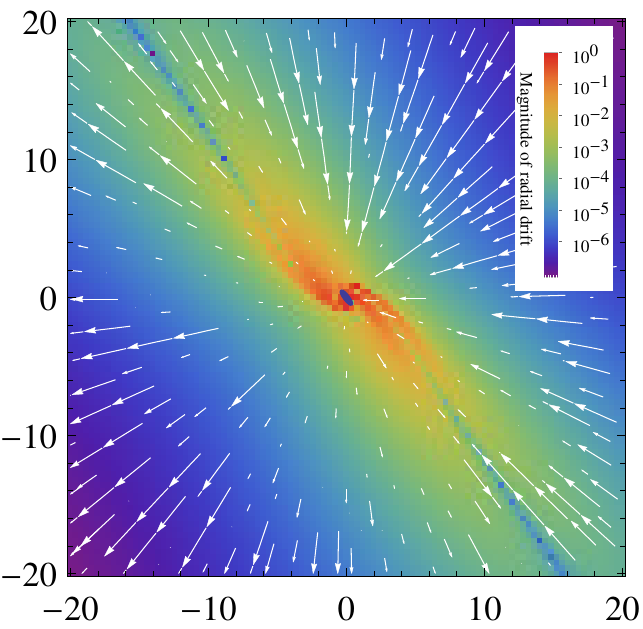}
\caption{The angular and the radial components of a type-0 `drift field'. The drift field is calculated by computing the displacement of a grid of individual trajectories after one period of the quantum state. Every individual coloured square depicted in the plots represents a data point on the $100\times100$ grid used. Such a drift field may be regarded as a time-independent velocity field of sorts, and may be used to track the long-term evolution of systems in regions where the velocity field has slow spatial variation. In this respect drift fields may be a useful tool for tracking the evolution of systems that are far from the bulk of the Born distribution (which may be considered to occupy a region in the centre of these plots with an approximate radius of 4). The arrows should be understood only to represent drift direction (the length of the arrows is meaningless). Instead, the wide variety in drift magnitude is represented by colour. The small elliptical orbit of the single node is indicated in navy blue. This is a type-0 drift field. In such fields, the predominant angular component is monotonic and the smaller radial component appears to produce equal inwards and outwards flow. A trajectory in such a drift field will circle the Born distribution with a mildly oscillating radius. In contrast to the type-1 and type-2 drift fields shown in figures \ref{fig2} and \ref{fig3}, such type-0 fields do not display a clear mechanism for a trajectory to approach the central region, a necessary precursor to quantum relaxation. Instead it could be the case that the trajectory does not approach the centre at all. If it eventually does reach the centre, it will certainly be significantly retarded with respect to type-1 and type-2 drift fields (examples in figures \ref{fig2} and \ref{fig3}). Hence, for states with type-0 drift fields, quantum relaxation of extreme quantum nonequilibrium may be frozen or in the least severely impeded.}\label{fig1}
\end{figure}

\begin{figure}
\includegraphics{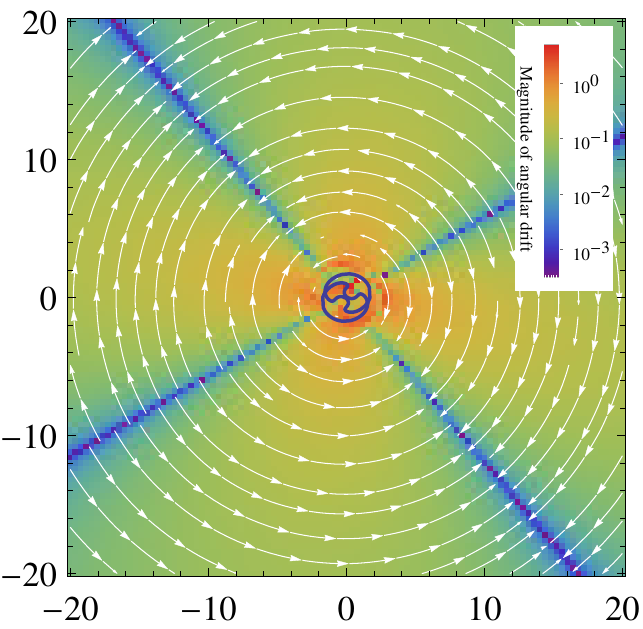}
\includegraphics{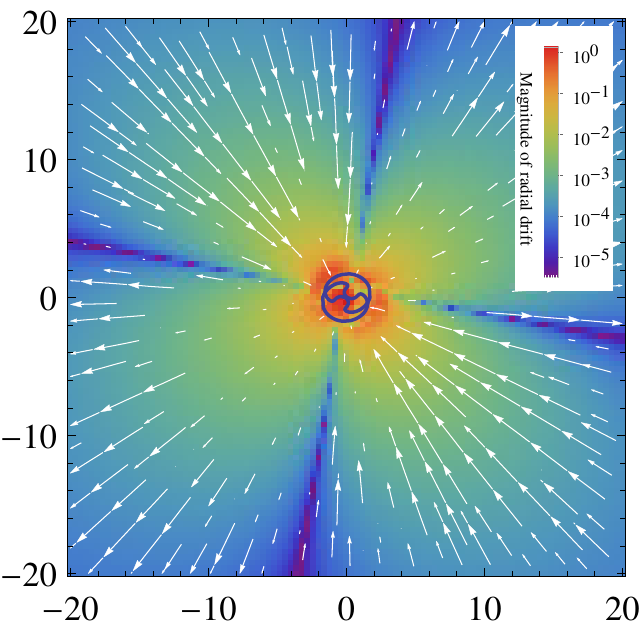}
\caption{An example of the angular and radial components of a type-1 drift field with the nodal paths indicated in navy blue. In type-1 drift fields the dominant angular flow features two axes that may or may not be perpendicular. One of these axes is repulsive and the other attractive. The radial component of the field may be variously divided into sections (slices) that are inwards towards the bulk of the Born distribution and outwards away from it. It always appears to be the case, however, that the axes that are angularly repulsive reside within the sections that are radially outwards, whilst the axes that are angularly attractive reside within the sections that are radially inwards. Such a structure provides a clear mechanism for the drift of systems into the bulk of the Born distribution. Firstly the predominant angular drift will draw any system towards the attractive axis, which for this quantum state runs from top-left to bottom-right. The system will become trapped on this axis, allowing the inwards radial drift to draw it into the central region. In this sense, type-1 drift fields possess a mechanism which enables relaxation even for distributions that are highly separated from quantum equilibrium.}\label{fig2}
\end{figure}

\begin{figure}
\includegraphics{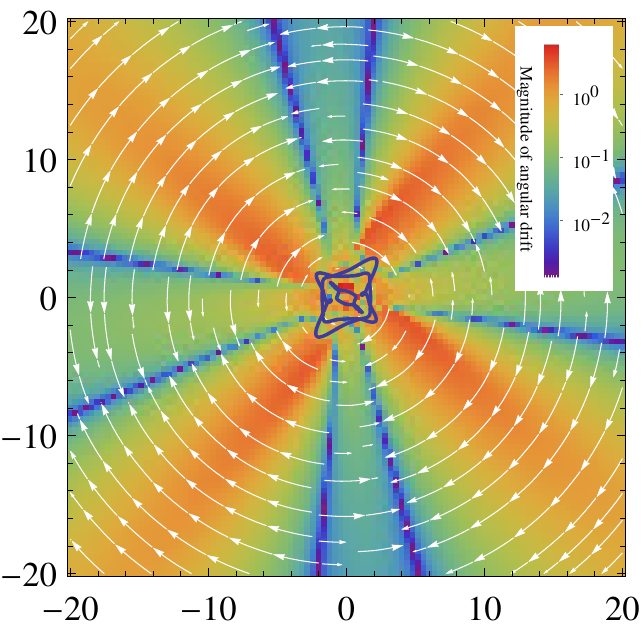}
\includegraphics{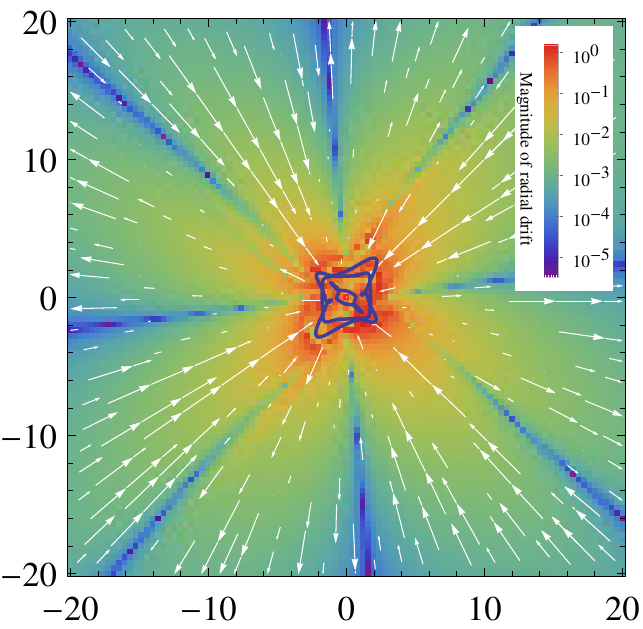}
\caption{An example of the angular and radial components of a type-2 drift field with the nodal paths indicated in navy blue. Such a field has a similar structure to a type-1 field except with two axes that are angularly attractive and two that are angularly repulsive. In the states studied, it is always the case that angularly repulsive axes align with regions with radially outwards drift, whilst angularly attractive axes align with regions with drift that is radially inwards. This provides a clear mechanism by which trajectories may be drawn into the central region (the Born distribution) regardless of where they are to be found initially. As such, for states with a type-2 drift field it is expected that even extreme forms of quantum nonequilibrium will relax efficiently.}\label{fig3}
\end{figure}

The chaos generated in trajectories close to the nodes should be understood to be in contrast to the regular behaviour of trajectories that are far from the nodes \cite{EKC07,ECT17}. Substitution of state \eqref{angular_expansion} and bases \eqref{angular_basis} into \eqref{angular_guidance} gives the guidance equations
\begin{align}
\overset{\circ}{\eta}&=\text{Im}\left\{\frac{\sum_{n_d n_g}C_{n_d n_g}\exp\left[-i(n_d+n_g)T+i(n_d-n_g)\varphi\right]\partial_\eta f_{n_d n_g}}{\sum_{n_d n_g}C_{n_d n_g}\exp\left[-i(n_d+n_g)T+i(n_d-n_g)\varphi\right] f_{n_d n_g}}\right\},\\
\overset{\circ}{\varphi}&=\frac{1}{\eta^2}\text{Im}\left\{\frac{\sum_{n_d n_g}i(n_d-n_g)C_{n_d n_g}\exp\left[-i(n_d+n_g)T+i(n_d-n_g)\varphi\right]f_{n_d n_g}}{\sum_{n_d n_g}C_{n_d n_g}\exp\left[-i(n_d+n_g)T+i(n_d-n_g)\varphi\right] f_{n_d n_g}}\right\}.
\end{align}
Accordingly, for large $\eta$ the physical velocity will scale as $\sim \eta^{-1}$ and will vary as $\sim \eta^{-2}$. Away from the bulk of the Born distribution and away from the nodes, one therefore generally expects to find a small and smooth velocity field.
This smoothness may be exploited as follows.
\begin{enumerate}
\item
 For a grid of initial positions, evolve trajectories with high precision through a single wave function period using a standard numerical method. (For this investigation a 5th order Runge-Kutta algorithm with Cash-Karp parameters was used.)
\item Record the final displacement vector from the original position for each grid point.
\item Plot the grid of displacement vectors as a vector field. This is done in figures \ref{fig1}, \ref{fig2} and \ref{fig3}. In these figures it was found to be useful to separate the angular and radial components of the field. This helps to distinguish the radial component from the generally large angular component.
\end{enumerate}
 The resulting `drift field' may be regarded as a time-independent velocity field of sorts, and used to track long-term evolution without the need for large computational resources as would usually be the case.
Of course, the drift field is merely intended to be indicative of long-term, slow drift and will certainly not be a good approximation in regions near to nodes where the motion is chaotic or otherwise quickly varying. Nevertheless one might suspect that, in regions where the field varies slowly with respect to the grid size upon which it is calculated, the drift field may capture the long-term evolution of the system well. 
The drift field has in practice proven very useful in classifying the global properties of long term evolution. 

As the production of such plots is relatively computationally inexpensive, it is possible to compute many such plots in a reasonable time. 
To study the typical behaviour of systems far out from the bulk of $|\psi|^2$ therefore, 400 drift field plots were calculated with 100 plots each for $M=3,6,10$ and 15 states. For the sake of comparison, superposed upon these plots were the numerically calculated trajectories of the corresponding nodes. The state parameters were randomised as follows. Random numbers in the range $[0,1]$ were assigned to all the $C_{n_d n_g}$ present in the state. These were then normalised with the factor $(\sum_{n_d n_g} C_{n_d n_g}^2)^{-1/2}$, before assigning a random complex phase. In all cases, the radial component of the drift field was notably smaller than the angular component, and in practice this can make the radial behaviour difficult to read from the plot. It was therefore useful to plot the radial component of the drift separately to the angular component. It was found that the angular and radial components of the drift fields displayed distinct global structures that could be readily categorized and which mirrored properties of the nodes of the state concerned. \\
\\
\textbf{Drift field structure and types}\\
In all the drift fields plotted, the drift was predominantly angular with only a small radial component. The structure of this dominant angular drift may be used to divide the drift fields into types 0, 1 and 2. Type-0 fields display flows that are entirely clockwise or entirely anti-clockwise. An example of a type-0 field is shown in figure \ref{fig1}. Type-1 fields, as shown in figure \ref{fig2}, feature one central attractive axis and one central repulsive axis which may or may not be perpendicular. Naturally, the prevailing flow is away from the repulsive axis and towards the attractive axis. Finally, type-2 fields are similar to type-1 except with additional axes. Specifically, type-2 fields feature two attractive axes and two repulsive axes. An example of a type-2 field is shown in figure \ref{fig3}. Considering the relative likelihood of each type of drift field for each of the states studied, it is likely that further types may appear for superpositions with components with larger cut-off energies ($m$ values) than studied here.

In contrast to the angular drift, the radial component of the drift field appears always to be (equally) divided into regions that are radially inwards and regions that are radially outwards. No state tested featured a radial field that was entirely radially inwards or entirely radially outwards. The simplest way in which the space may be divided is by a single dividing axis through the origin as shown in figure \ref{fig1}. In this case it is evident that there is as much of the space with outwards drift as there is with inwards drift. Often it is the case that two or more of these axes divide the space (see figures \ref{fig2} and \ref{fig3}), so that the space is apportioned into wedges that alternate between flow that is radially inwards and flow that is radially outwards. In the case of two such axes (as for instance is shown in figure \ref{fig2}), it appears that these axes are always perpendicular, so that again half of the space exhibits inwards flow and the other half outwards flow. More complicated, bulb-like structures may sometimes be seen in the radial drift, especially in larger superpositions. To the eye however, it always appears the case that the space is \emph{equally} divided into inwards and outwards regions, although a mathematical proof of whether this is indeed the case remains to be seen.\\
\\
\textbf{Mechanism for the relaxation of extreme quantum nonequilibrium}\\
Any nonequilibrium distribution that is initially located away from the central region (where the Born distribution is located) cannot be considered relaxed until the vast majority of systems have at least relocated into the central region. 
As discussed in the introduction, it is expected that once this happens relaxation will proceed efficiently.  
It therefore becomes relevant to consider the time it takes an individual trajectory to migrate into the central region (or indeed whether this migration takes place at all).
To this end, type-1 and type-2 drift fields have a clear, identifiable mechanism by which this migration takes place.
Type-0 drift fields do not feature this mechanism. In every type-1 and type-2 field calculated, the angularly repulsive axes appear to coincide with portions of the field with inwards radial drift whilst the angularly attractive axes coincide with regions that are radially inwards. This structure is clearly displayed in figures \ref{fig2} and \ref{fig3}. In this regard, any system that is initially separated from the bulk of the Born distribution will be swept by the dominant angular drift towards one of the angularly attractive axes. As these axes always coincide with a region of inwards radial drift, the trajectory will then be dragged into the central region where the bulk of the Born distribution is found.

In contrast to this, type-0 drift fields produce trajectories that perpetually orbit around the central region. As a trajectory does so it will sample both regions (wedges) with inwards drift and regions with outwards drift. As these appear (at least to the eye) equal in size it is tempting to conclude that, for such trajectories, outwards drift will balance inwards drift. If this were the case then trajectories would not be drawn into the central region and relaxation would not take place. Certainly it is the case that there is a balancing effect between inwards and outwards flow and so for type-0 fields relaxation from outside the central region will be at least significantly retarded if not stopped altogether. Whether or not there exists some delicate imbalance between inwards and outwards flow that eventually produces relaxation remains to be seen. An answer to this question could have important implications for studies considering conjectured quantum nonequilibrium in the early universe, and so this is clearly a direction for further work.
\begin{figure}
\includegraphics{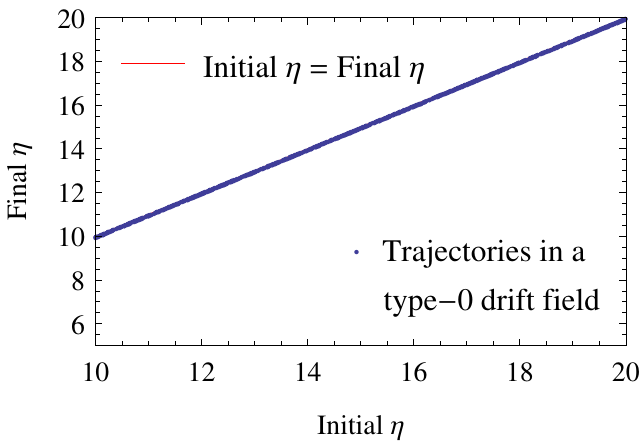}
\includegraphics{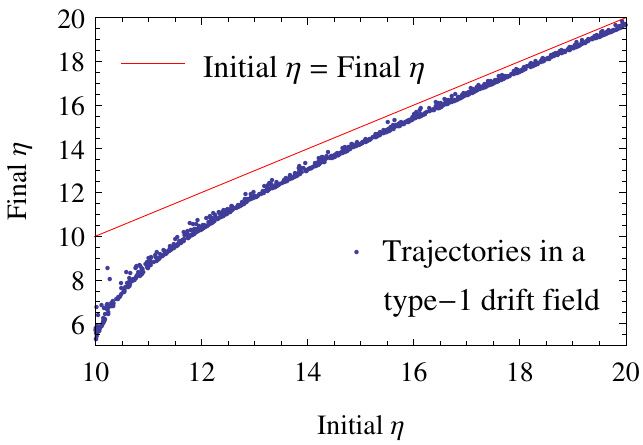}
\caption{A comparison of radial displacement of trajectories after 1000 periods produced by the type-0 and type-1 drift fields shown in figures \ref{fig1} and \ref{fig2}. The type-1 drift field causes a clear net drift inwards that appears to be absent in the type-0 field. Under the influence of a type-1 drift field, an `extreme' nonequilibrium distribution (that is concentrated away from the central region--small $\eta$) will be transported to the bulk of the Born distribution where it is presumed that it will relax efficiently to quantum nonequilibrium. In contrast, the balancing effect on radial drift for type-0 fields (discussed in figure \ref{fig1}) results in no such clear radial displacement. These frames were produced by placing 1000 trajectories randomly in the interval $10<\eta<20$ and numerically evolving for 1000 wave function periods with a standard Runge-Kutta algorithm. }\label{fig4}
\end{figure}

Although a detailed study is left for future work, the effect of the relaxation mechanism (the resulting radial transport of trajectories) is illustrated in figure \ref{fig4}. To produce this figure, 1000 trajectories were chosen randomly in the radial interval $10<\eta<20$, and numerically evolved for 1000 wave function periods with a standard Runge-Kutta algorithm. In the first frame of the figure, the wave function parameters that produced the type-0 drift field displayed in figure \ref{fig1} were used. In the second frame the wave function parameters that produced the type-1 drift field of figure \ref{fig2} were used. In accordance with our statements regarding the mechanism, the type-1 drift field causes a clear inward drift of the trajectories. An `extreme' quantum nonequilibrium ensemble (that is concentrated away from the central region) will therefore be transported towards the Born distribution where, as discussed, it is presumed to relax efficiently. In contrast, the type-0 drift field produces no such radial drift and, after 1000 wave function periods, the trajectories still appear close to their original radial coordinate $\eta$.  \\
\\
\textbf{Specific properties of states studied}\\
\textbf{M=3}\\
An $M=3$ state has only $\chi_{00}$, $\chi_{10}$ and $\chi_{01}$ components. (Or equivalently $\psi_{00}$, $\psi_{10}$ and $\psi_{01}$ components.) Since the cut-off energy is $m=1$, by property (xi) the total vorticity of the state may only be $\mathcal{V}_\text{tot}=\pm2\pi$. Hence, there must always exist a single node with $\mathcal{V}=\mathcal{V}_\text{tot}$. Property (xii) ensures there is a maximum of one node and so no pair creation can take place. (The two plane algebraic curves mentioned at the end of section \ref{sec:nodes} are in this case straight lines and so intersect only once.) The sign of the vorticity may be determined with the vorticity theorem. The relevant Laurent polynomial $C_{01}/z+C_{10}z$ has a simple pole at $z=0$ and two simple zeros at $z=\pm\sqrt{-C_{01}/C_{10}}$. Hence, if $|C_{10}|>|C_{01}|$ the vorticity is positive, else it is negative. Clearly the two possibilities appear with equal probability when wave function parameters are chosen at random. For the 100 states generated, the drift field was type-0 in every case. Hence, the relaxation of extreme nonequilibrium may be retarded for every $M=3$ state. The angular flow is in the direction indicated by $\mathcal{V}_\text{tot}$. Note however that due to the way the $\chi_{n_d n_g}$ are defined, $d$ should be understood to refer to positive, anticlockwise vorticity rather than the common notion of rotating to the right implied by the letter $d$.

For the simple $M=3$ case, further properties of the node may be found as follows. By taking the modulus square of the state \eqref{xy_state}, the path of the node may be written $AQ_x^2+BQ_xQ_y+CQ_y^2+DQ_x+EQ_y+F=0$ (which is the general Cartesian form of a conic-section) with the coefficients
\begin{align}
A=d_{10}^2,\enspace B=(D_{10}D^*_{01}+D^*_{10}D_{01}),\enspace C=d_{01}^2,\enspace D=E=0,\enspace F=-\frac12 d_{00}^2.
\end{align}
The condition that this path is elliptical is $B^2-4AC<0$, which is the case unless $\theta_{10}$ and $\theta_{01}$ differ by an exact integer number of $\pi$, the limiting case in which the ellipse becomes a straight line. 
That the ellipse is centred upon the origin is implied by the vanishing $D$ and $E$ coefficients.
The nodal trajectory is circular if $B=0$ and $A=C$, which is the case when $\theta_{01}$ and $\theta_{10}$ differ by a half-integer number of $\pi$ and $d_{10}=d_{01}$, or equivalently $D_{10}=\pm i D_{01}$. By translating the angular basis, it may be shown that the ellipse has a semi-minor axis of $c_{00}/(c_{10}+c_{01})$ and a semi-major axis of $c_{00}/|c_{10}-c_{01}|$. (A circular trajectory is found if either $c_{10}$ or $c_{01}$ vanish.) As the closest the node approaches the origin is $c_{00}/(c_{10}+c_{01})$, for a ground state with only perturbative contributions from $\psi_{10}$ and $\psi_{01}$ (as studied by \cite{KV16}), the node stays far from the bulk of the Born distribution and interesting relaxation properties are to be expected. The orientation of the ellipse varies linearly with the difference between $\phi_{01}$ and $\phi_{10}$. The area of the ellipse is $\pi c_{00}^2/|c_{10}^2-c_{01}^2|$. 
The nodal trajectory as a function of time may be expressed
\begin{align}
Q_x(T)=\frac{1}{\sqrt{2}}\frac{d_{00}}{d_{10}}\frac{\sin(\theta_{01}-\theta_{00}-T)}{\sin(\theta_{10}-\theta_{01})},\quad
Q_y(T)=\frac{1}{\sqrt{2}}\frac{d_{00}}{d_{01}}\frac{\sin(\theta_{10}-\theta_{00}-T)}{\sin(\theta_{01}-\theta_{10})}.
\end{align}
\textbf{M=6}\\
By property (xi) the total vorticity may only be $-4\pi$, $0$ or $4\pi$. As an $M=6$ state is bounded in energy by $m=2$, by property (xii) there is a maximum of 4 nodes at any time. In the case of $\mathcal{V}_\text{tot}=\pm 4\pi$ vorticity, there must always exist at least 2 nodes, each with vorticity $\mathcal{V}=\mathcal{V}_\text{tot}/2$, and one pair may create and annihilate in addition to these. All states generated with $\mathcal{V}_\text{tot}=\pm 4\pi$ produced type-0 drift fields. In the case of $\mathcal{V}_\text{tot}=0$, there may be zero nodes, or up to two pairs of opposite vorticity nodes. (The plane algebraic curves upon whose intersections the nodes reside are in this case conic sections which may intersect 0, 2 or 4 times.) The Vorticity 0 states produced type-1 drift fields. If the wave function parameters are chosen at random in the way described, the $\mathcal{V}_\text{tot}=0$ type-1 drift fields appear approximately 66\% of the time whilst the $\mathcal{V}_\text{tot}=\pm 4\pi$ type 0 appear approximately 17\% of the time each. (The relative frequency of states was determined by randomly selecting state parameters for 100,000 states and calculating $\mathcal{V}_\text{tot}$ using the vorticity theorem \eqref{polynomial}.) Hence, states that exhibit retarded relaxation are relatively less common than for $M=3$ states, appearing in 34\% of cases.\\
\\
\textbf{M=10}\\
By property (xi) the total vorticity may be $\pm 2\pi$ or $\pm6\pi$, with the former possibility always featuring at least one node and the latter at least 3 nodes. In both cases property (xii), allows pair creation to increase this number up to a total of 9 nodes. The $\mathcal{V}_\text{tot}=\pm6\pi$ are relatively rare, each appearing in approximately 3\% of randomly selected cases each. All of the 6 randomly generated states that had $\mathcal{V}_\text{tot}=\pm6\pi$ produced type-0 drift fields. States with $\mathcal{V}_\text{tot}=\pm2\pi$ each appear in approximately 47\% of cases. These may exhibit any one of the three types of drift field. Of the 94 fields generated by states with $|\mathcal{V}_\text{tot}|=2\pi$, type-0 drift fields (with retarded relaxation) appeared in 7 cases, whilst type-1 and type-2 fields appeared in 72 and 15 cases respectively.\\
\\
\textbf{M=15}\\
By property (xi) the total vorticity may be 0, $\pm4\pi$ or $\pm8\pi$. States with $\mathcal{V}_\text{tot}=0$ may feature no nodes, but states with $\mathcal{V}_\text{tot}=4\pi$ and $\mathcal{V}_\text{tot}=8\pi$ must always retain at least 2 and 4 nodes respectively. For all possible vorticities, property (xii) and pair creation allow up to a maximum of 25 nodes. As with the $M=10$ superpositions, the states of maximal total vorticity are relatively rare. For $M=15$ superpositions, maximal vorticities $\mathcal{V}_\text{tot}=\pm8\pi$ each appear in approximately 0.4\% of cases, whilst $\mathcal{V}_\text{tot}=\pm4\pi$ each appear in approximately 19\% of cases. The most commonly found vorticity is $\mathcal{V}_\text{tot}=0$, appearing in approximately 61\% of cases. The single maximal vorticity $\mathcal{V}_\text{tot}=\pm8\pi$ state that appeared in the 100 randomly selected states produced a type-0 retarded drift field. Of the 42 random trials that had $|\mathcal{V}_\text{tot}|=4\pi$, 15 had type-0 drift fields. Of the remaining 57 trials that were found to have $\mathcal{V}_\text{tot}=0$, not a single one produced a type-0 drift field.\\
\\
\textbf{Vorticity-drift conjectures}\\
Although proofs have not been forthcoming, we advance the following statements as conjectures and argue their validity primarily on the basis of lack of counterexample.\\
\\
\textbf{Conjecture 1 - A state with zero total vorticity cannot produce a type-0 drift field}\\
For sufficiently large $\eta$, the velocity field of a state with zero total vorticity cannot be uniformly clockwise or uniformly anticlockwise. 
To demonstrate this, consider that for a zero-vorticity state it is the case that $\int_{0}^{2\pi}\frac{\partial S}{\partial\varphi}\mathrm{d}\varphi=0$ for sufficiently large $\eta$. Hence, the integrand $\partial S/\partial\varphi$ (which is proportional to the angular component of the velocity field) must change sign or be trivially zero (in the presence of fine tuning). If, then, the drift field classifications were applied instead to velocity fields, it would be the case that zero total vorticity fields could not produce type-0 velocity fields. To support the validity of this statement when applied to drift fields, 1000 extra zero-vorticity states were generated for each of the $M=6$ and $M=15$ categories. (Property (xi) means $M=3$ and $M=10$ states cannot have zero-vorticity.) (States with maximal or zero total vorticity may be easily calculated by randomly generating many states and checking the vorticity with the vorticity theorem.) All 2000 cases were found to be type-1 or type-2, in support of the conjecture.\\
\\
\textbf{Conjecture 2 - All states of maximal vorticity produce type-0 drift fields}\\
Consistently it was found that all states of maximal vorticity produced type-0 relaxation retarding states. As these states are relatively rare for $M=10$ and $M=15$ categories, however, this conjecture is certainly in need of further substantiation. To provide this, 1000 extra maximal vorticity states were generated for each of the $M=3,6,10,15$ state categories.  All 4000 cases were found to be of type-0. We note that this conjecture provides a convenient method of generating these type-0 relaxation retarding states were one to wish to study their properties. One need simply to randomly generate state parameters and then check that their vorticity is maximal using the vorticity theorem.

\section{Conclusions}
\label{sec:conclusion}
In this paper, consideration has been given to the behaviour of de Broglie trajectories that are separated from the bulk of the Born distribution with a view to describing the quantum relaxation properties of more `extreme' forms of quantum nonequilibrium. The main results are as follows. 
A number of new properties of nodes have been described ((vii) to (xii) in section \ref{sec:nodes}) that are true under the assumption of a quantum state with bounded energy and in the absence of finely-tuned parameters. It is hoped that these results prove useful to the community and that they may be extended and generalised to suit problems other than that considered here. For the 2-dimensional isotropic oscillator (which has physical significance in studies regarding quantum nonequilibrium in the early universe \cite{AV10,CV13,CV15,CV16,UV15,UV16}), these properties have been shown to have consequences for the quantum relaxation of systems that are separated from the bulk of the Born distribution (what has been referred to as `extreme' quantum nonequilibrium). The relaxation properties of these systems have been classified in terms of the structure of their `drift field', a new concept introduced here. States have been divided into 3 distinct classes. Type-1 and type-2 drift fields have been shown to feature a mechanism that gives rise to efficient relaxation of extreme quantum nonequilibrium. Type-0 fields lack such a relaxation mechanism, appearing instead to sample the drift field in a manner that suggests the prevention of quantum relaxation. Even if it is not the case that relaxation is entirely blocked in such states, it will at least be significantly retarded. Whether or not some delicate imbalance of flow may eventually cause relaxation, and the calculation of relaxation timescales if this is indeed the case, presents a clear avenue for further investigation. Another subject that will be the focus of a future paper concerns the consequences of fine-tuning and perturbations around finely-tuned states. (In the process of this investigation, such states were found to produce highly unusual, pathological behaviour that was judged to be extraneous to the intended focus.)

 For the states studied, it was found that all states of maximal total vorticity $\mathcal{V}_\text{tot}$ (according to their energy bound) produced type-0, relaxation-retarding drift fields. It was also found that no zero-vorticity states produced exhibit type-0 drift. In section \ref{sec:drift} it was formally conjectured that these two correspondences are true for all (non-finely-tuned) states. These conjectures are a central result of this work which, when used with the vorticity theorem (point (x) in section \ref{sec:nodes}), allow the generation of states that should exhibit efficient relaxation as well as those for which relaxation is retarded (or possibly stopped altogether). The relative abundances of these states given random parameters have been discussed. It is hoped that the identification of relaxation retarding states, and the methodology that has otherwise been formulated, may be useful to those investigating the intriguing possibility of discovering quantum nonequilibrium. 

\ack
I would like to thank Lucien Hardy for his invitation to visit Perimeter Institute, and also Antony Valentini for his supervision and many useful discussions.
This research was supported in part by Perimeter Institute for Theoretical Physics.  Research at Perimeter Institute is supported by the Government of Canada through the Department of Innovation, Science and Economic Development Canada and by the Province of Ontario through the Ministry of Research, Innovation and Science.
\section*{Bibliography}
\bibliography{/home/nick/Research_x230t/citations}
\end{document}